\documentclass{tJOT2e}

\usepackage{graphicx}% Include figure files
\usepackage{dcolumn}% Align table columns on decimal point
\usepackage{bm}% bold math
\jvol{00}
\jnum{00}
\jmonth{00}
\jyear{00}

\def\Bd#1{{\it {\bf #1}}}

\def\B.#1{{\it {\bf #1}}}

\def\i{{\rm i}}

\def\bse{\begin{subequations}}
\def\ese{\end{subequations}}
\def\Re{{\rm Re}}

\def\alpha{k}
\def\s{\left(k_+^2 + \beta^2\right)}
\def\A{\left[\i (\omega_+ - k_+ U) + {1 \over R}(D^2 - k_+^2 - \beta^2)\right]}

\def\2d{two-dimensional }
\def\3d{three-dimensional }

\def\jfm{J. Fluid Mech. }
\begin{document}

%\preprint{APS/123-QED}

\title{The signature of laminar instabilities in the zone of 
transition to turbulence}

\author{N. Vinod \& Rama Govindarajan \break Engineering Mechanics Unit,
\break Jawaharlal Nehru Centre for Advanced Scientific Research, \break Jakkur, 
Bangalore 560 064 INDIA \break e-mail: nvinod@jncasr.ac.in, rama@jncasr.ac.in }

\received{\today}

\maketitle
\begin{abstract}

We demonstrate that the space-time statistics of the birth of turbulent 
spots in boundary layers can be 
reconstructed qualitatively from the average behavior of macroscopic 
measures in the transition zone. The conclusion in \cite{vg04} that 
there exists a connection between the patterns in laminar instability 
and the birth of turbulent spots is strengthened.
We examine why the relationship between instability and 
transition to turbulence is manifest in some cases and appears to be 
totally absent in others. 
Novel cellular automaton type simulations of the transition zone are 
conducted, and the pattern of spot birth is obtained from secondary 
instability analysis. 
The validity of the hypothesis of concentrated breakdown, according 
to which most
turbulent spots originate at a particular streamwise location, is
assessed.
The predictions made lend themselves to straightforward experimental 
verification.
\end{abstract}

%\pacs{47.27.Cn, 47.27.Nz }

\section{Introduction}

\begin{figure}[h]
\begin{center}
\includegraphics[scale=0.5]{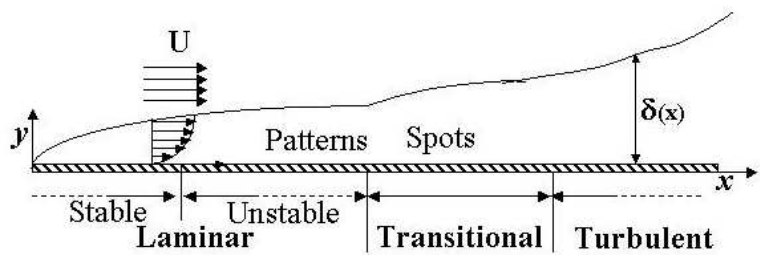}
\end{center}
\caption{Sequence of events in the laminar-turbulent transition 
process, 
on a boundary layer formed by the flow past a semi-infinite plate. }
\label{fig:bl}
\end{figure}

The physics of the transition to turbulence in a boundary layer has
been the subject of a large number of theoretical and 
experimental investigations. While most of the process is understood,
as shown in the schematic picture in figure \ref{fig:bl},
there are still gaps in our knowledge, as we shall discuss below. 
The process begins with the linear amplification of two-dimensional 
disturbance waves of a narrow band of frequencies. Once the 
disturbance modes have grown to a certain amplitude, they 
destabilize three-dimensional secondary modes 
\cite{henni98,saric96,saric03}. These modes display a locally regular 
pattern of maxima, which may be aligned (harmonic) or staggered 
(subharmonic). Further downstream, in what is termed the transition 
zone, turbulent spots, or concentrated patches of turbulence, are
found, surrounded by quiet laminar flow. The spots grow as the
convect downstream, and merge with each other to make the flow
asymptotically fully turbulent. Our interest here is in the breakdown 
of instability waves into turbulent spots. We do not explain the 
complex physics occurring between the zone of secondary instability and
the region containing spots, but we present evidence showing that the 
pattern existing in the first is retained for the most part in the 
second.

It is at present almost impossible either experimentally or 
numerically to track the birth and downstream growth of a large 
number of individual turbulent spots and to obtain detailed 
statistics. It is however not difficult to measure average quantities 
such as turbulent intermittency (the fraction of time that a flow is 
turbulent). With these, and knowing the behaviour of a typical spot, we
demonstrate that it is possible to reconstruct the scenario of spot 
birth. It was recently shown \cite{vg04} that intermittency behaviour 
in many flows is consistent with spot birth as dictated by the most
unstable secondary mode. We present another example of this in 
decelerating flows.
Hitherto, it was assumed that spot birth is completely random. 
No contradiction to this assumption was presented by intermittency 
measurements in low external noise environments, in the flow past a 
flat plate aligned with the flow. Since most intermittency measurements
were made for this specific flow, the assumption of random spot birth
gained credence and was not re-examined. We show that in this special
case, the intermittency variation obtained from both random and regular
spot birth are similar, and discuss why. We then present other
measures which even in this case would differentiate clearly between
random and regular spot birth. 
We exploit the versatility of our simulation method to demonstrate
(section \ref{sec:axisym}) that the effect of geometry (in the form 
of transverse curvature) is significant in the boundary layers on 
cylinders.
It is hoped that the present 
predictions will motivate experimental efforts.
 
We conduct secondary instability computations in the standard manner
(section \ref{secondary}), and employ novel cellular automaton type 
stochastic simulations for spot birth, growth and convection, as 
described in section \ref{sec:trans}. We make a number of assumptions. First,
while the secondary instability analysis is three-dimensional, the
stochastic simulations are conducted in a plane parallel to the solid
body, when viewed from above. This means that the variations within
the boundary layer in a direction $y$ normal to the wall are averaged over.
The $x$-$y$ profile of turbulent spots is not uniform,
but given that they are flat objects and the boundary layer thickness is
very small ($O(R^{-1})$ compared to the other dimensions), averaging over $y$
is a valid approximation. 
Secondly, we employ the hypothesis of concentrated breakdown, 
according to which all spots are born at a particular streamwise 
location \cite{57Nara}, i.e. at a particular Reynolds number.
We show in section \ref{sec:conc} that this is a reasonable assumption.
We also make use of the experimentally and numerically observed shape
and downstream behaviour of single spots, especially the fact that spot growth 
is self-similar at any pressure gradient.

A spatial pattern in turbulent in the form of spirals and spots
are observed in the experiments 
of Prigent {\em et al.} \cite{prigent} in plane Couette flow
and Taylor-Couette flow. 
They found that the spatial modulation of turbulent intensity
obeys the dynamics of coupled amplitude (Ginzburg-Landau type) 
equations with noise. It is important to note that the
simulations of \cite{jd01,wu} in boundary layers establish that structures in the unstable
laminar region are the precursors of turbulent spots. These simulations
were of bypass transition. While the present work is about the
traditional Tollmien-Schlichting mechanism taking place in
low-disturbance environments, the basic connection between instability
and transition is similar.

\section{Analysis of secondary instability}
\label{secondary}

The growth of small (linear) perturbations in a boundary layer, 
under the assumption of parallel flow,
is described by the Orr-Sommerfeld equation \cite{orr,sommer}
\begin{equation}
(U-\omega/\alpha)(v''-\alpha^2v)-U''v=\frac{1}{\i\alpha R}
(v^{\rm iv}-2\alpha^2v'' +\alpha^4v),
\label{os}
\end{equation}
where $U(y)$ is the mean velocity and $R$ the Reynolds number, 
$\equiv U_\infty(x) \theta(x)/\nu$, $U_\infty$ is the local
freestream velocity, $\theta$ is a characteristic boundary layer
thickness, and $\nu$ the kinematic viscosity. $v(y)$ is the 
amplitude of the normal component $\hat v$ 
of the disturbance velocity, given by
$$
\hat v(x,y,t) = v(y) \exp[\i(k x - \omega t)],   
$$
$k$ and $\omega$ being the wavenumber and the frequency of the
disturbance mode being considered. 
The primes denote differentiation with respect to $y$. 

When the linear modes have grown to a significant amplitude
(of the order of a percent of the mean flow), the new
flow becomes unstable to secondary, typically three-dimensional, 
disturbances.
The linearised equations describing the secondary instability are 
obtained by writing flow quantities in the form \cite{88Herb,pre03}
%$$
\begin{equation}
\Bd U(x,y,t) = \Bigg\{U(y) \vec i + A\left[u(y) \vec i + v(y) \vec
j\right]
%$$
%{\rm \quad \quad \quad}
\exp
\left[\i \left(\alpha x - \omega t\right)\right]\Bigg\}
+\vec {\Bd u_s(x,y,z,t)}.
\label{basic}
\end{equation}
The quantity in the curly brackets is the basic flow consisting 
of the laminar profile plus the primary instability wave. The primary
instability is very slowly growing, and the flow may be taken to be
periodic. $A$
stands for the amplitude of the primary disturbance.
The three-dimensional secondary disturbance is written as
%$$
\begin{equation}
\Bd u_s(y,\Bd r_\perp,t)=
 \Re\Big\{\Bd u_{s+}(y)
\exp\left[\i\left(\Bd k_+\cdot
\Bd r_\perp - \omega_+ t\right)\right] 
%$$
%{\rm \quad \quad \quad}
+ \Bd u_{s-}(y)
\exp\left[\i\left(\Bd k_-\cdot
\Bd r_\perp - \omega_- t\right)\right]\Big\},
\end{equation}
where $\Bd r_\perp\equiv x\vec i+z\vec k$ and
$\Bd k_\pm\equiv k_\pm \vec i \pm \beta \vec k.$
We substitute the above ansatz into the Navier-Stokes and continuity 
equations, retain linear terms in the secondary, and eliminate the 
disturbance
pressure and streamwise component of the velocity. On averaging over
$x$, $z$ and $t$, only the resonant modes survive, which are related as
follows:
\begin{equation}
k_+ + k_- = \alpha, \qquad {\rm and} \qquad (\omega_{+} + 
\omega_-)_r = \omega.
\end{equation}
We then obtain
$$
\A \left[\s f_{s+} - Dv_{s+}\right] - \i k_+ U' v_{s+} -
$$
$$
A_p\left({k_+ \over 2 k_-}\right)\Bigg\{\left[\i k_+ u D + v D^2 +
\i k_- Du \right]v_{s-}^*
+ \bigg[\left(\beta^2-k_-k_+\right)vD + 
$$
\begin{equation}
\i k_+ \left(k_-^2+\beta^2\right)u\bigg]f_{s-}^*\Bigg\} = 0,
\label{first}
\end{equation}
$$
{\rm and} \quad \A \left(Df_{s+} - v_{s+}\right) 
- \i k_+ U' f_{s+}
$$
$$
 + A\left({\alpha + k_- \over 2}\right) \left[{v \over k_-}Dv_{s-}^* -
\i u \left(v_{s-}^* + Df_{s-}^*\right)\right]
$$
\begin{equation}
+{A \over 2}\left[v\left({\alpha\beta^2 \over k_-} + D^2\right) - \i k_-(Du)
\right]f_{s-}^* = 0.
\label{second}
\end{equation}

The quantity $f_{s+}(\equiv - \i w_{s+}/\beta)$ is proportional to 
the spanwise component of
the secondary disturbance velocity, and the operator $D$ stands for
differentiation with respect to  $y$.
The boundary conditions are
\begin{equation}
\Bd u_s=0 \quad {\rm at} \quad y = 0 \quad {\rm \&} \quad
\Bd u_s \rightarrow 0 \quad {\rm as} \quad y \rightarrow \infty.
\label{bcs}
\end{equation}
Equations (\ref{first}) and (\ref{second}), along with two corresponding
equations in $v_{s-}^*$ and $f_-^*$, describe an eigenvalue problem for 
the secondary instability. Note that secondary instabilities account for
one contribution to nonlinearity, that of the triad interaction between 
the primary Tollmien-Schlichting mode and two three-dimensional 
disturbances, the sum of
whose wave-numbers equals that of the primary. 

A Chebyshev spectral collocation method is used for the numerical 
solution \cite{canuto}. A grid stretching given by 
\begin{equation}
y_j={b(1+Y_j) \over 1 + {2 b / y_\infty} - Y_j}
\end{equation}
is applied, where 
\begin{equation}
Y_j=\cos({\pi j \over N_g}), \qquad j=0,1,2,...,N_g,
\end{equation}
are Chebyshev collocation points. We thus transform the 
computational domain from $(-1,1)$ to $(0,y_\infty)$, where $y_\infty$ 
is chosen to be at least $5$ times the boundary layer thickness, and 
cluster grid points close to the wall by tuning the parameter $b$. 
Results for the boundary layer over a flat-plate compare very well
with those of \cite{88Herb}.
On changing the number of grid points $N_g$ from $80$ to $160$, the 
eigenvalues remained identical up to the sixth decimal place.

We are interested here in the flow past flat solid surfaces inclined 
at an angle $2m/(m+1)$ to the flow, where the 
external flow velocity varies downstream as $U_\infty \sim x^m$ 
\cite{batchelor}. We consider constant pressure ($m=0$) and 
decelerating boundary layers ($m < 0$) in this paper, 
since the behaviour in the transition zone differs in the two. 
The profile of mean streamwise velocity for a given $m$ is described by 
the Falkner-Skan equation
\cite{fmwhite}
\begin{equation}
f'''+ff''+ \frac{2m}{m+1}(1-f'^2)=0
\end{equation}
where $U=f'$. The boundary conditions are $f=f'=0$ at the wall, and
$f' \to 1$ as $y \to \infty$. 
Computations of instability have been carried out for a variety of Reynolds 
number and primary disturbance amplitude. 
The sub-harmonic mode, of wavenumber $(=k_+=k_-)=\alpha/2$ is found
always to be the most unstable.

\section{The transition zone}
\label{sec:trans}

The transition zone is easiest to describe quantitatively in terms of
the variation with the streamwise coordinate, $x$ of the intermittency,
$\gamma$. Given that the Reynolds number $R(x)$ is the only parameter
in the instability problem, it is reasonable to assume that most
turbulent spots will originate within a narrow spanwise strip around
a particular streamwise location $x_t$, denoted here as the location of
transition onset. This is the hypothesis of concentrated breakdown 
\cite{57Nara}. With spots appearing randomly in accordance with a 
Poisson distribution in time and a uniform distribution in the spanwise 
coordinate $z$, $\gamma$ would vary downstream as \cite{57Nara}
\begin{equation}
\gamma =1- \exp\left[{-n\sigma \over U_\infty} (x-x_t)^2\right],
\end{equation}
or
\begin{equation}
F\equiv \sqrt{-\log(1-\gamma)} =\sqrt {-n\sigma \over U_\infty} (x-x_t)
\label{gamma}
\end{equation}
such that the intermittency parameter $F$ varies linearly in $x$.
Here $U_\infty$ is the free stream velocity, $n$ is the number of spots forming
per unit length in the spanwise direction, $z$, per unit
time, at $x_t$. The shape and downstream growth of turbulent
spots is experimentally observed to be similar in all pressure gradients:
a turbulent spot maintains an arrowhead shape when viewed from above, 
and remains self-similar as it grows
\cite{85Nara,95sw}. It is convenient therefore to define a
non-dimensional spot propagation parameter as
\begin{equation}
\sigma=\left[{1 \over U_r}-{1 \over U_h}\right]U_\infty \tan \zeta,
\label{sigma}
\end{equation}
where $\zeta$ is the angle subtended by the spot at its origin and
$U_h$ and $U_r$ respectively are the speeds with which its head and rear 
convect. 

It is relatively straightforward to measure other average transition 
zone quantities such as the burst rate and the persistence time 
distribution of laminar or turbulent flow as functions of the 
streamwise distance. The burst rate $B$ is here defined as 
the mean number of switches from laminar to turbulent flow per 
unit time at a given location, and the
persistence time $W$ of laminar flow is the duration of time for which
the flow remains continuously laminar at a given location. The
transitional intermittency has been measured in many experiments, but there
are no data available for the other measures, as far as we know.
It is our suggestion that a lot of information can be gained from
knowing how these quantities vary in the streamwise direction.
For a Poisson process, the burst rate is related to the local 
intermittency by
\begin{equation}
B \propto (1-\gamma)\left(-\ln(1-\gamma)\right)^{1/2},
\label{burst}
\end{equation}
and the probability density function of persistence time $W$ of laminar flow
is
\begin{equation}
p(W) \propto (1-\gamma)^{W-1}\left(-\ln(1-\gamma)\right).
\label{persist_lam}
\end{equation} 

\begin{figure}[h]
\begin{center}
\vskip0.5in
\includegraphics[scale=0.3]{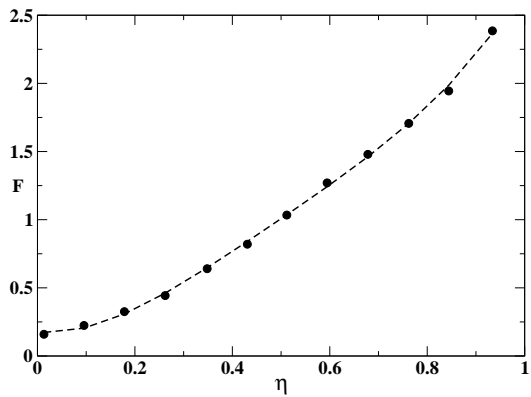}
\vskip0.3in
\caption{Variation of the intermittency parameter $F$ in adverse pressure 
gradient, $m=-0.05$. Symbols: experiment \cite{88gb}.}
\label{fig:F0b}
\end{center}
\end{figure}

In quiet facilities, the parameter $F$ in equation \ref{gamma} in constant 
pressure flow (to be discussed later) varies almost perfectly linearly 
with $x$, but its 
variation in decelerating flows is nonlinear with an increasing slope, 
as seen in figure \ref{fig:F0b}. 
The explanation \cite{vg04} for the nonlinear behaviour of 
intermittency is that spot birth is not random, but mainly regular, in
a pattern dictated by the most unstable secondary mode.
\vskip2mm 

\section{Stochastic Simulations}

\begin{figure}
\begin{center}
\includegraphics[scale=0.4]{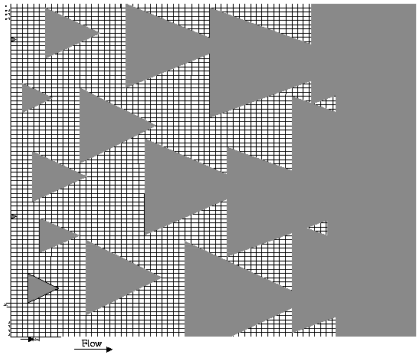}
\caption{Schematic diagram of the simulation domain in a random spot breakdown
scenario. Spots appear according to a Poisson distribution in time, and are
uniformly distributed in the spanwise direction.}
\vskip3mm
\label{fig:randx}
\end{center}
\end{figure}

\begin{figure}
\begin{center}
\includegraphics[scale=0.4]{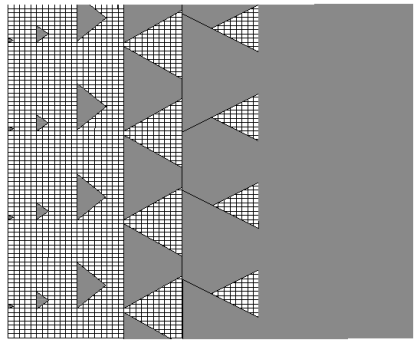}
\caption {Schematic diagram of the simulation domain in regular breakdown. 
Spots are formed at regular intervals in the spanwise direction and are
staggered after every time period, in a manner dictated by the secondary 
instability.}
\vskip3mm
\label{fig:reg1}
\end{center}
\end{figure}

Stochastic simulations of the transition zone (using a cellular 
automaton-like approach) are performed, employing observed properties 
of spot growth and convection \cite{vg04,vinod_pramana}. We consider two scenarios 
for spot birth. The first is a random breakdown, as discussed above,
with no relation to instability. The second is predominantly regular 
breakdown, where the pattern prescribed is that of the maxima in
disturbance vorticity in the secondary instability.
Simulated transition zones for random and regular (harmonic) 
breakdown are shown schematically figures \ref{fig:randx} and 
\ref{fig:reg1} respectively.

The zone, as viewed from above, is discretized into $200 \times 400$ 
rectangular grids in $x$ and $z$.
Each grid is assigned an integer $\chi$, 
equal to $0$ if the flow there is laminar and $1$ if it is turbulent. 
A regular breakdown scenario is prescribed as follows.
During the first time interval, one spot is generated every
$N_z$ sites, at $X=1$ and $Z=l N_z+i$, $l=0,1,2 \cdots L_z/N_z$,
where $i$ is an integer between $1$ and $N_z$. The uppercase stands for
discretised coordinates.
After $N_t$ time steps the spots form at spanwise locations which
are staggered in the spanwise coordinate with respect to the spots
generated earlier,
i.e., at $X=1$ and $Z=l N_z+j$ for harmonic and
$Z=(l+1/2) N_z+j$ for subharmonic breakdown
where $j=i+1$.
This prescription corresponds to spots forming at
the crests of an oblique wave of spanwise and streamwise wavenumbers
\begin{equation}
\beta = {2\pi \over N_z\Delta Z}
{\rm \qquad and \qquad}
\alpha = {2\pi \over N_t\Delta T v}
\label{discbeta}
\end{equation}
respectively, where $v$ is the streamwise velocity of the wave crest.
The phase of the
oblique wave is randomised with a small probability ($1\%$), and a small
fraction ($5\%$) of the spots are generated randomly. It is assumed
that this prescription mimics qualitatively the randomisation
due to external disturbance. Changes in these fractions do not change the 
answers significantly.

In a random breakdown, spots appear at $X=1$ in accordance with a Poisson distribution 
in time and a uniform distribution in the spanwise direction. 
In each case, the downstream growth of turbulent spots follows the same 
algorithm. During one time interval, the front of each spot moves ahead 
by $2$ grid locations, while the rear moves forward by $1$ grid location. 
The lateral dimension of the spot is increased by $1$ grid location on 
either side. The simulated spot is thus triangular, and retains its shape
for all time. A real spot is a blunter at its leading edge than a 
triangle, but this is assumed not to change the results significantly.
 
In experiments, the speed of the front of the spot is comparable to the 
flow velocity $U_\infty$ while the rear moves forward approximately at $0.5U_\infty$. 
The half-angle $\zeta$ subtended by the spot at its origin
is about $10^\circ$ in a constant pressure boundary layer, and increases with 
adverse pressure gradient. 

\begin{figure}
\begin{center}
\includegraphics[scale=0.30]{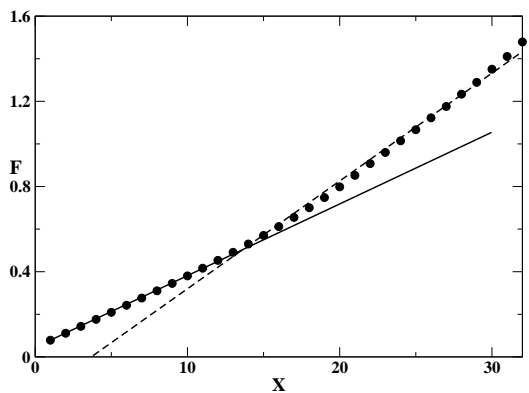}
\caption {Streamwise variation of the intermittency parameter for $N_z=49$
and $N_t=10$. Symbols: regular breakdown,
solid line: random breakdown. The mean spot generation rate is kept constant
at $N=0.01$. The dashed line shows the best linear fit for the downstream
part of $F$.}
\label{fig:fv}
\end{center}
\end{figure}

\begin{figure}
\begin{center}
\includegraphics[scale=0.30]{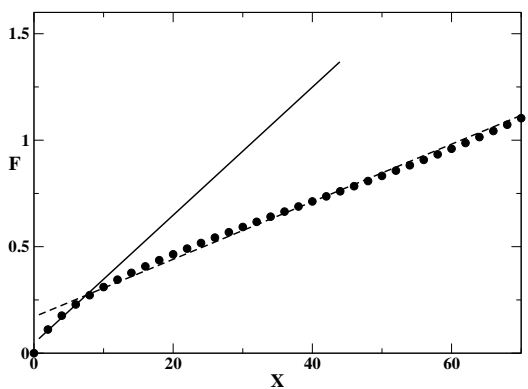}
\caption {Streamwise variation of the intermittency parameter for $N_z=10$ 
and $N_t=49$. Other parameters and notation are the same as in figure 
\ref{fig:fv}. It can be seen that the transition zone is much longer in this 
case. }
\label{fig:fw}
\end{center}
\end{figure}

In Figs. \ref{fig:fv} and  \ref{fig:fw}
it is demonstrated that intermittency behaviour is sensitive to the ratio
of spanwise to streamwise wavenumbers. This observation is used
to strengthen the connection we make between instability and its
signature in the transition zone.
The simulations shown in Fig. \ref{fig:fv} display a distinct change in slope
at $X=13$, just downstream of the location ($X=11$) at which the
heads of a set of spots formed at a time $T$ just touch the rears of the
spots formed at $T-N_t\Delta T$.
In the simulations of Fig. \ref{fig:fw}, all parameters have been kept the same,
except that the values of $N_z$ and $N_t$ have been switched. In this case
the spots touch each other laterally first (at $X=6$) downstream of which
a sharp change in slope is again evident. At low $F$, i.e., upstream of spot
merger, the intermittency behaviour in each case is exactly the same as that when
spots form randomly (with the same breakdown rate). This
is not surprising, given that the spots are too small to ``see'' each other.
At higher intermittencies, quite surprisingly, for the same mean breakdown rate
and identical spot growth, the length of the transition zone is seen to be very
different, which tells us that spot-merger plays a large role in transition.
A purely lateral merger results in a much higher degree of overlap (on both 
sides), and consequently, less of the region is occupied by spots. This of 
course means that transition to turbulence will proceed much more slowly as 
shown in Fig. \ref{fig:fw}. A purely longitudinal merger on the other hand, 
results in far less spot overlap, meaning more occupancy by spots, and a 
shortened transition zone (Fig. \ref{fig:fv}). Incidentally, it is often the
case that experimental results seem to lie on two straight lines, as in this
figure. This has been termed a  `sub-transition' \cite{rn84}. Our simulations
show that this change in slope in not a result of spots suddenly changing
their downstream growth behaviour, but is a natural consequence of their
spatial arrangement.

We now ask which kind of merger scenario would be consistent with
growing instability modes. To do this, we perform a secondary instability 
analysis to obtain the streamwise and spanwise wavenumbers for which the
secondary wave grows fastest. For example, for an
adverse pressure gradient flow of Falkner-Skan parameter $m=-0.06$,
the maximum
growth rate ($\omega_i$) of the secondary instability wave is $0.067$, and occurs
when the wavenumbers ($k$ and $\beta$) are $0.165$ and $0.215$ respectively.
The corresponding phase speed is $0.524 U_\infty$.
Using the fact that the rear of a spot travels at approximately half the freestream
velocity, we have $\tan\zeta = 2 \Delta Z/U_\infty\Delta T$. The spot propagation
angle is taken as $\approx 20^o$, which is close to the experimental value 
for this pressure gradient \cite{95gmw}, and from equation (\ref{discbeta})
we may estimate the breakdown ratio $N_z /N_t$ to be $ \sim 4.9$.
This ratio changes only by $2\%$ for a $50\%$ change in Reynolds number, and
may be taken to be independent of Reynolds number.

\begin{figure}
\begin{center}
\includegraphics[scale=0.30]{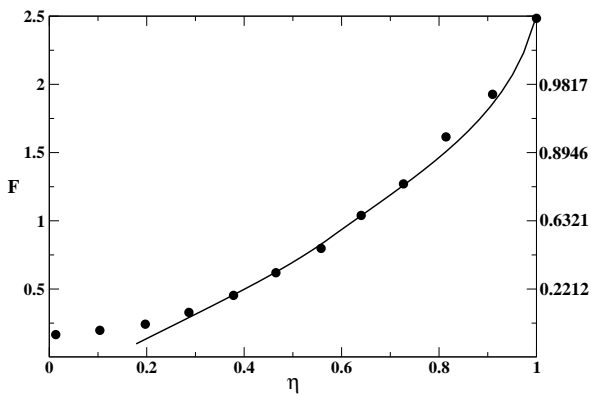}
%\caption {The parameter $F$ corresponding to the intermittency shown in 
%figure \ref{fig:gam}}.
\caption{Downstream variation of the intermittency parameter $F$
in an adverse pressure gradient of $m=-0.05$.
Solid line: stochastic simulations with $N_z=39, N_t=12$, from
the dominant secondary mode, symbols: experiments \cite{94gbw}.
}

\label{fig:grow1}
\end{center}
\end{figure}

In figure \ref{fig:grow1} we consider an adverse pressure 
gradient, $m=-0.05$. The ratio $N_z$/$N_t$ obtained from stability analysis 
for this case is $3.23$, and in the stochastic simulations we use 
$N_z=39$ and $N_t=12$. 
The analysis is done at a Reynolds number of $220$, based on the boundary 
layer momentum thickness, to match the experimental value.
A similar result for $m=-0.06$ is available in \cite{vg04}. 
The results of both simulations are in good agreement with experiments.
In particular, the changing slope of the intermittency
parameter $F(x)$ is followed well. 
We
may conclude that a spot breakdown pattern as prescribed by
the secondary instability gives rise to intermittency behaviour in qualitative
agreement with  measurements in highly decelerating flows.

%\begin{figure}
%\begin{center}
%\includegraphics[scale=0.30]{g_0410n.eps}
%\caption{Downstream variation of the intermittency
%in an adverse pressure gradient of $m=-0.05$.
%Solid line: stochastic simulations with $N_z=39, N_t=12$, from
%the dominant secondary mode, symbols: experiments \cite{94gbw}. 
%}
%\vskip3mm
%\label{fig:gam}
%\end{center}
%\end{figure}

It may be noticed that there is some discrepancy at the beginning 
of the transition zone, where it is known that experimental data may be
inaccurate: errors could arise unless the data is collected over extremely long
times. Secondly, tiny patches of turbulence can be difficult to
distinguish from noise. It is also to be remembered that the hypothesis of
concentrated breakdown is an idealisation: in reality spots would be
forming within a narrow streamwise strip around $x=x_t$, rather than at one
particular location, this can smear out the intermittency at $\gamma \sim 0$.

\begin{figure}
\begin{center}
\includegraphics[scale=0.45]{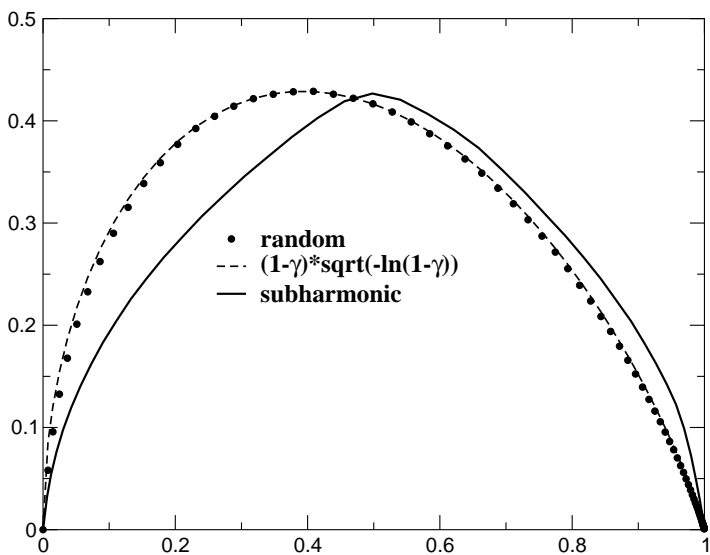}
\caption{
Variation of the burst rate with intermittency. The solid line shows a
subharmonic breakdown, with $N_z/N_t=4.9$, with $5\%$ randomness. The
scale on the ordinate is arbitrary. The result is similar to the
predominantly regular harmonic breakdown case shown in \cite{vg04}. The
dashed line shows equation \ref{burst} and the circles are
simulation results with random breakdown.
}
\label{fig:burst}
\end{center}
\end{figure}

The burst rates obtained from the present simulations with a random breakdown 
are in agreement with equation \ref{burst}, as seen in figure 
\ref{fig:burst}. Any deviation from this 
behaviour is a sign that spot
breakdown is not random. When the breakdown is mostly regular, it is clear
that the burst rate distribution is ``peaky" and symmetric with respect
to $\gamma=0.5$. The burst rate variation
with intermittency is similar for sub-harmonic and harmonic type breakdowns.
We suggest here that there is a need for experimental measures of these
quantities, especially in adverse
pressure gradient boundary layers, which will help us quantify the breakdown
scenario. The burst rate behaviour displayed when the dominant merger is 
lateral is qualitatively different from the case where the merger is 
longitudinal. The burst rates obtained from the simulations whose 
intermittencies are shown in figure \ref{fig:fv} and \ref{fig:fw}, are plotted
in figure \ref{fig:burst}, and are a demonstration of this. 
 
%\begin{figure}[h]
%\begin{center}
%\includegraphics[scale=0.33]{bg_lat.eps}
%\caption{Variation of the burst rate with the intermittency for different 
%scenarios of spot birth. The regular pattern conforms to sub-harmonic.}
%\label{fig:burst_lateral}
%\end{center}
%\end{figure}

\begin{figure}[h]
\begin{center}
\includegraphics[scale=0.33]{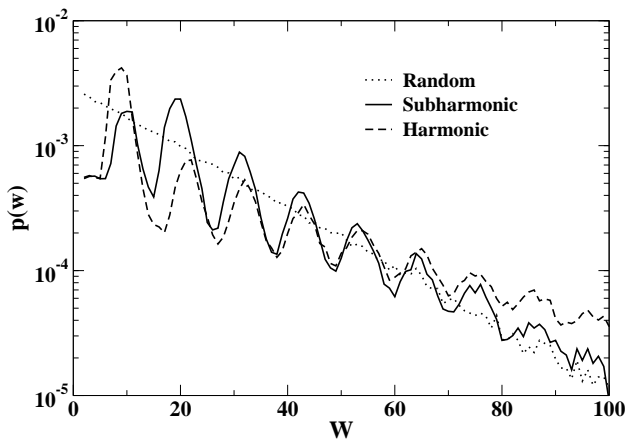}
%\vskip 5mm
%\includegraphics[scale=0.33]{w_g50.eps}
\caption{Probability density function of persistence time of laminar flow at (a)
$\gamma=0.1$. Dashed line: random breakdown, solid line: 
$90\%$ regular (sub-harmonic), long dashes: $50\%$ regular.  
}
\label{fig:per_stag}
\end{center}
\end{figure}

\begin{figure}[h]
\begin{center}
\includegraphics[scale=0.33]{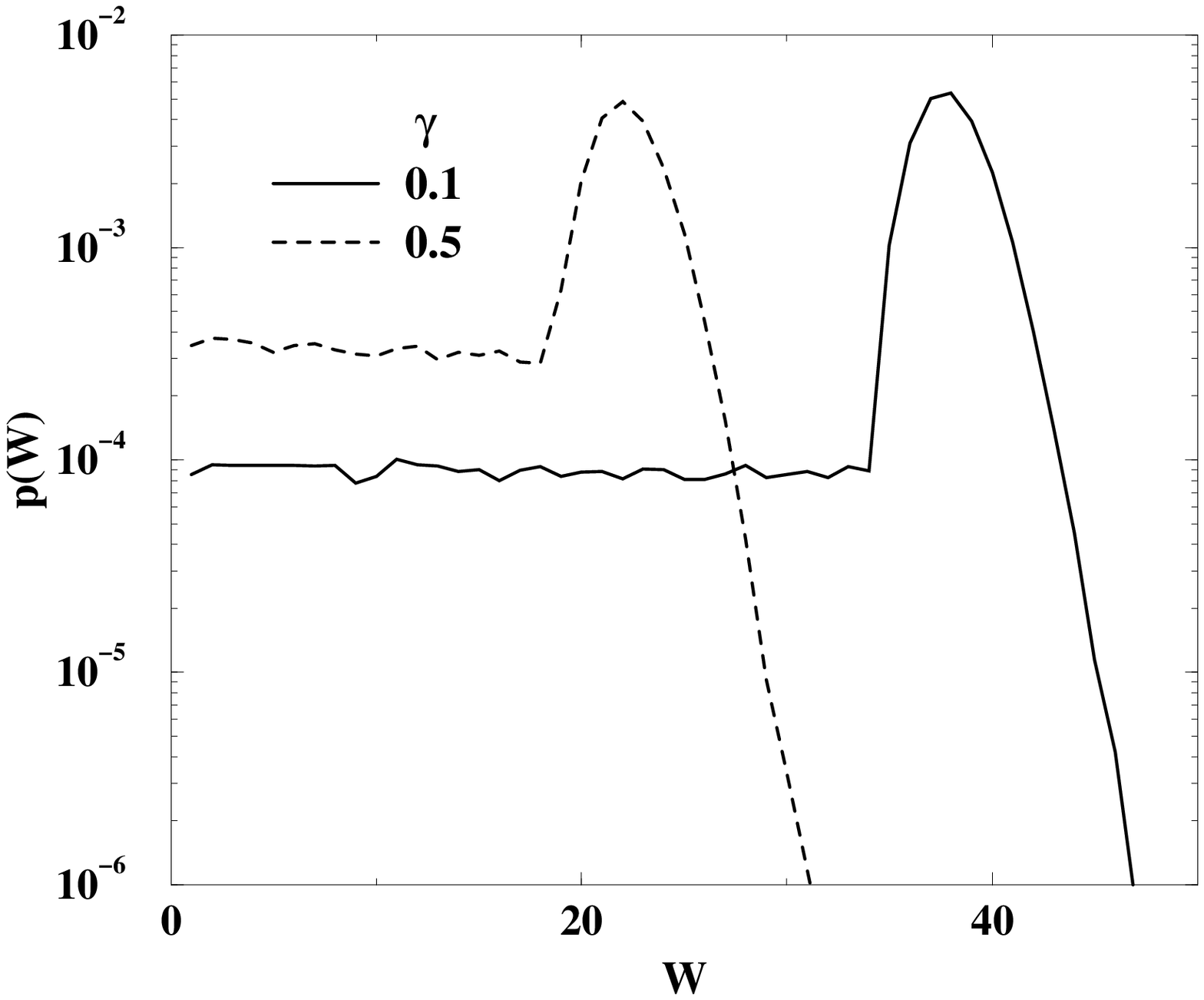}
\caption{Probability density function of persistence time of laminar flow at 
$\gamma=0.1$ (solid line) and  $\gamma=0.5$ (dashed line), The simulations
are carried out under conditions where lateral merger is dominant as in 
Fig. \ref{fig:fw}.}
%%$90\%$ regular (sub-harmonic T2), long dashes: $50\%$ regular. Thin 
%%dot-dashed line: predominant lateral merger as in Fig. \ref{fig:fw}.}
\vskip3mm
\label{fig:per_lat}
\end{center}
\end{figure}

The probability density function of the persistence time $W$
is plotted in figure \ref{fig:per_stag}. The data is obtained by running the
simulation over 20 million time steps (after reaching stationary state),
monitoring a particular streamwise location in the middle of the span
and collecting statistics of lengths of strings of zeroes between two 1's. 
The reverse information, about strings of 1's between two zeroes, is
less interesting.
We present results for $W$ at location where the intermittency is
$\gamma=0.1$. For a random spot breakdown, the probability density 
of the persistence time decays exponentially with persistence time, as
is expected from equation (\ref{persist_lam}).
At low intermittency levels, 
in the case of regular breakdown, an overall decay is significantly
modulated by ups and downs. At higher levels of intermittency, if the breakdown 
is predominantly regular, the probability of very large waiting times is 
extremely low, as expected. The most probable persistence times correspond 
to the modulated streamwise extent of the laminar zone between two rows of spots.
In the case of subharmonic (staggered) spot arrangement, there is a 
probability of skipping one row of spots, giving larger persistence times. 
In figure \ref{fig:per_lat} the probability density function of the 
persistence time from a simulation where the lateral merger is dominant is
plotted. The qualitative behaviour is different and an experimental
measurement of $p(W)$ is can thus indicate the dominant breakdown pattern. 

\section{Is spot birth random in constant pressure gradient flows?}
\label{sec:zero}
 
\begin{figure}
\begin{center}
\includegraphics[scale=0.30]{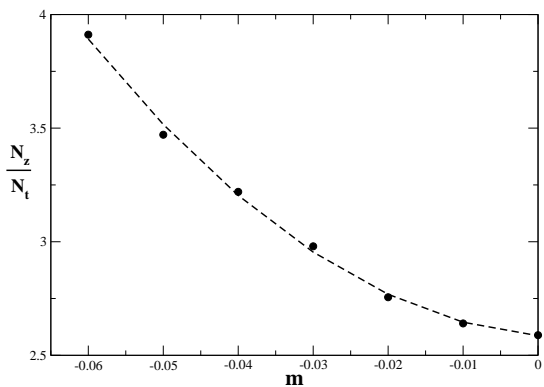}
\caption {The variation with pressure gradient of the ratio $N_z/N_t$ as 
determined by the most unstable secondary mode.  }
\label{fig:ratio}
\end{center}
\end{figure}

\begin{figure}
\vskip7mm
\begin{center}
\includegraphics[scale=0.30]{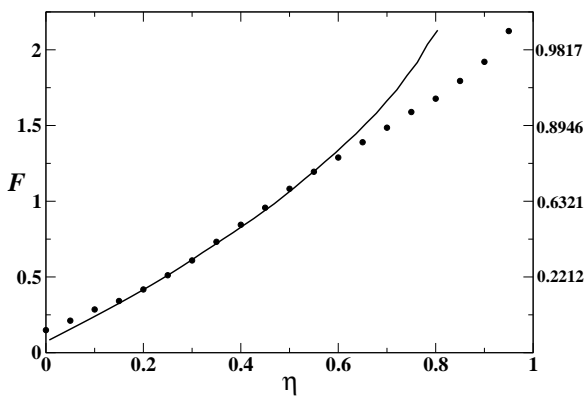}
\caption
{
Variation of intermittency parameter $F$ in the boundary layer over a flat
plate. Solid line: stochastic simulations with $N_z=32, N_t=13$ as
dictated by secondary instability, the dots are experimental data in constant
pressure boundary layers \cite{dhawan58}.
}
\label{fig:zero}
\end{center}
\vskip3mm
\end{figure}

In constant pressure (flat plate boundary layer) flow, two kinds of
behaviour are observed. In high-disturbance environments, the
intermittency behaviour is similar to that seen in adverse pressure
gradients, except the dominant mechanism is the secondary instability
of streaks. The experimentally measured intermittency in such a flow
\cite{matsubara} has been shown \cite{vg04} to be consistent with the 
dominant instability pattern. Most experiments, however, are conducted
in flat plate in environments which are painstakingly maintained quiet.
Here, the intermittency parameter $F$ is linear in $x$, consistent
with random spot-birth. Do  secondary instabilities not directly
trigger spot birth here? A partial answer may be gleaned from 
figure \ref{fig:ratio}, where the spanwise/ streamwise ratio of wavelengths 
of the dominant mode is shown as a function of the pressure gradient.
The ratio decreases with
decreasing pressure gradient, and $N_z /N_t$ is about $2.3$ for a constant 
pressure
boundary layer (since now $k=0.085$, $\beta=0.146$ and $v=0.353$ for the most
dangerous mode). For a regular breakdown with this ratio, simulations show
that the variation in $F$ is linear up to about
$\gamma=0.75$ (figure \ref{fig:zero}), the slope increases towards the end of the transition zone.
On the lines of the discussion on Fig. \ref{fig:fv} and \ref{fig:fw},
 we may expect that spot mergers are now a
combination of lateral and longitudinal, with longitudinal being marginally
dominant. The weakened dominance means that the slope change occurs much closer 
to full turbulence, at $\gamma \sim 0.8$. 

The obliqueness of the most unstable wave provides only a partial 
answer for why the variation of $F$ is linear in a constant pressure boundary
layer. The perfectly linear fit in a vast number of experimental
data is also likely to be because the degree of randomness in this case is actually
higher than in an adverse pressure gradient boundary layer, as suggested
by Gostelow and Walker \cite{90gw}. The following qualitative
arguments indicate why we may expect this.

\begin{figure}
\begin{center}
\includegraphics[scale=0.4]{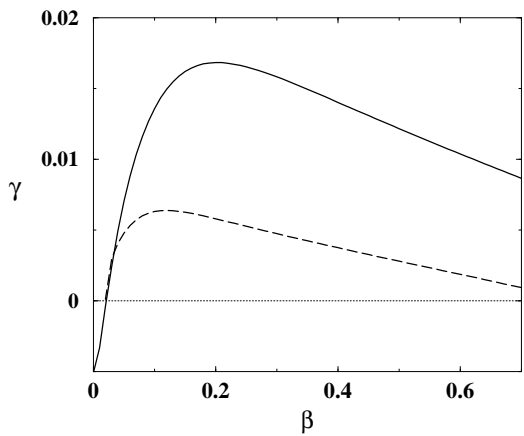}
\caption{Growth rate of most dangerous secondary sub-harmonic mode at a
Reynolds number $R=600$, based on the momentum thickness of the boundary
layer and $U_\infty$. Solid line: $m=-0.06, k_+=0.14$, dashed line:
$m=0, k_+=0.085$.}
\vskip3mm
\label{fig:omi}
\end{center}
\end{figure}

\begin{figure}
\vskip3mm
\begin{center}
\includegraphics[scale=0.3]{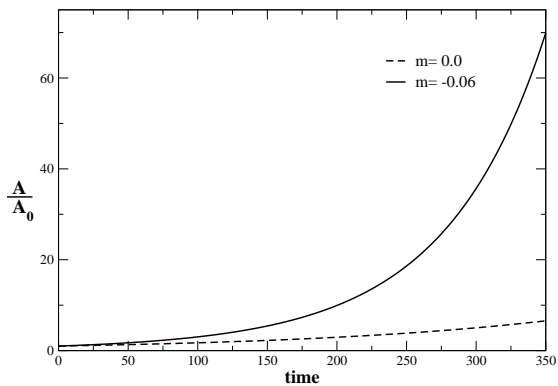}
\caption{
The variation of amplitude of secondary disturbance wave with time in adverse
($m=-0.06$, $\alpha=0.185$ and $\beta=0.12$,solid line) and zero 
($\alpha=0.13$ and $\beta=0.805$, dashed line) pressure gradient boundary 
layers.
}
\label{fig:wave}
\end{center}
\end{figure}

A decelerating flow is inherently much more unstable \cite{89co,99masl}, as 
demonstrated in figures \ref{fig:omi} and \ref{fig:wave}. Since the primary wave
is slowly growing, we assume that the secondary mode at a given time can be
computed by taking the primary wave to be of constant amplitude (this is a
temporal analog of the ``parallel-flow" assumption often employed in
spatially-developing flow).
It is to be noted that the growth of the secondary mode can be
faster than exponential if the primary mode is unstable as well.
Within a very short downstream distance, the disturbance waves
in an adverse pressure gradient flow achieve the threshold amplitude 
required for breaking down into spots. In the constant pressure case
the attainment of the required threshold is much slower,
offering greater opportunity for stochastic effects.
 
\begin{center}
\begin{table*}
\begin{tabular}{|c|c|c|c|}
\hline
Pressure gradient ($m$) & Reynolds number & phase speed & group velocity \\
\hline
  0     &  200   &  0.412  & 0.295 \\
 -0.05   &  200   &  0.492  & 0.655 \\
  0     &  600   &  0.353  & 0.188 \\
 -0.05   &  600   &  0.510  & 0.666 \\
\hline
\end{tabular}
\caption{Typical phase and group velocities in zero pressure-gradient and
decelerating boundary layers}
\vskip4mm
\end{table*}
\end{center}

\begin{center}
\begin{table*}
\begin{tabular}{|c|c|c|c|} \hline
\multicolumn{2}{|c|}{ } & \multicolumn{2}{|c|}{$k_+/\beta$} \\ \hline
$A$ &$R$ & $m$ =0.00 & $m$=-0.049\\
\hline
0.01 & 200 & 0.77 & 0.75 \\
0.005 & 200 & 0.91 & 0.70 \\ \hline
0.01 & 600 & 0.65 & 0.70 \\
0.005 & 600 & 0.81 & 0.75 \\ \hline
\hline
\end{tabular}
\caption{Effect of the primary disturbance amplitude on the obliqueness
of the most dangerous mode.}
\end{table*}
\vskip2mm
\end{center}
 
A significant amplitude modulation of wave packets could be another cause 
for randomness in zero pressure gradient flow.
Amplitude modulation would result in different waves reaching the threshold
at different streamwise stations, and spot birth would be smeared out
over a streamwise distance. In the case of decelerating flow,
the group velocities of the waves are much larger, as seen in Table I, and the
amplitude modulation takes place over a much narrower width.
At a Reynolds number of $200$, the group velocity $c_g$ of the most unstable
secondary in adverse pressure gradient is more than twice the corresponding wave
over a flat plate. At a Reynolds number of $600$, this ratio is more than $3$.
The phase speeds on the other hand are not that different.
 
Finally, the wavelength
ratio of the dominant mode varies significantly with primary disturbance
amplitude in flat-plate flow (see \cite{93zel}). This, however, is not the case in
strongly decelerating flow, as seen in Table II. If the external noise is 
irregular in amplitude, a greater randomisation would take place in flat 
plate flow than in decelerating flow.

These results indicate that in a decelerating boundary layer, 
formed by the flow past an inclined plate, the connection between 
instability and transition is likely to be much easier to observe, and we 
therefore recommend experimental work in adverse pressure gradient, rather 
than on flat plate boundary layers.

\section{Effect of concentrated breakdown}
\label{sec:conc}

\begin{figure}
\begin{center}
\includegraphics[scale=0.30]{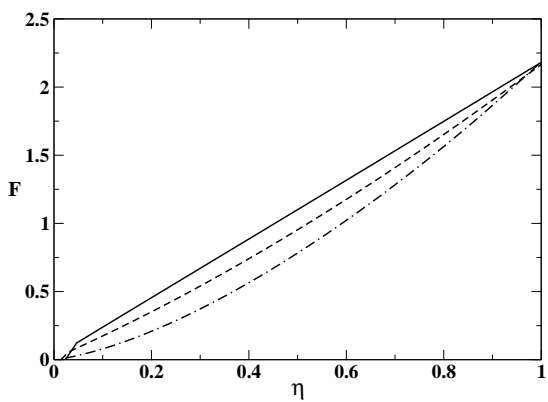}
\caption{Effect of the hypothesis of concentrated breakdown on the
intermittency distribution.
Solid line: spots are allowed to form only at $x_t$.
Dashed line: $80\%$ spots are born downstream of the onset location.
Dot-dashed line: all spots form downstream of the transition onset (Emmons
breakdown). }
\label{cb:f}
\end{center}
\end{figure}

\begin{figure}
\begin{center}
\includegraphics[scale=0.3]{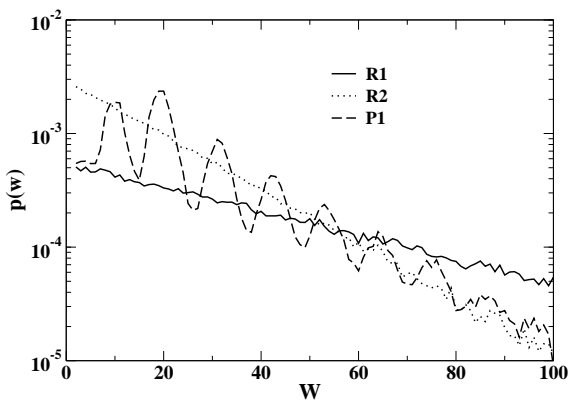}
\caption{
Effect of concentrated breakdown on persistence time distribution.
The persistence time is computed at location where intermittency $\gamma=0.1$.
The curve marked R2 is according to the hypothesis of concentrated breakdown, 
R1 is for random breakdown anywhere downstream of $x_t$ \cite{51emm}, 
P1 represents periodic breakdown (sub-harmonic),
 the pattern is obtained from secondary instability. }
\label{cb:wait}
\end{center}
\end{figure}

Narasimha \cite{57Nara} proposed the hypothesis of concentrated breakdown
in which all spots form within a narrow spanwise belt around the location of 
transition onset, i.e., within $x_t \pm \epsilon$, where $\epsilon << x_t$.
While the intermittency behaviour resulting from the hypothesis and the 
assumption of a random breakdown matches experiment very well in constant
pressure and low disturbance environments, its validity has been a matter
for debate, see e.g. \cite{johnson94}. 
Since a particular disturbance amplitude is achieved at a particular 
Reynolds number, it is plausible at least that upstream of a given $x$ 
location, no spots will form. A large number of spots are likely to form
in the vicinity of $x_t$, but given that instability modes which are not 
dominant could continue to grow in the transition zone, a small number 
could be born at any location downstream. 
We conduct simulations to estimate 
how much the intermittency and other parameters depend on making this
hypothesis. 
We do this by allowing an increasing fraction of the spots
to form with equal probability anywhere downstream of $x_t$. A random
breakdown is prescribed. Our conclusion is that even in
the unlikely event of a large fraction of spots being born downstream,
the intermittency is dominated by spots forming at $x_t$, since at any 
given $x$, these spots are much larger than those born more recently.
The intermittency distributions for the extreme cases tried are shown in 
figure \ref{cb:f}. It is seen that even when $80\%$ spots form downstream of
the onset location, the variation of $F$ is still practically linear.
However when all the spots are born downstream of $x_t$ 
(as modelled by Emmons \cite{51emm}), there is a noticeable departure from 
linearity. In an instability-driven transition, it is highly likely, as 
discussed above, that a significant fraction of spots will be born around 
$x_t$. It may be noticed that the intermittency behaviour in the case of 
Emmons breakdown does not appear very different qualitatively from a regular 
breakdown, and would thus be difficult to distinguish in an experiment. 
However, an experiment which measures persistence times can distinguish 
very simply between them, as is evident from figure \ref{cb:wait}. In an 
Emmons breakdown, the persistence time distribution decays exponentially, 
but with a smaller slope than for the concentrated breakdown, as expected. 

\vskip4mm

\section{Axisymmetric Boundary Layers}
\label{sec:axisym}
Transition to turbulence in the boundary layers forming around axisymmetric 
bodies is poorly understood in
spite of wide application in the motion of submarines, fishes etc. 
When the transverse curvature is significant, transition can proceed quite 
differently from a two-dimensional boundary layer \cite{rn84,gn00}. When the
typical patch of turbulence, consisting
either of a single spot or a group of spots which have merged laterally,
attains a width of the order of the diameter of the cylinder, it wraps
itself around the body. Downstream of the location of wrap,
further lateral growth is not possible. The turbulent patch
then resembles a sleeve \cite{gnv74} displaying only a one-dimensional
growth in the streamwise direction.

\begin{figure}[h]
\begin{center}
\includegraphics[scale=0.5]{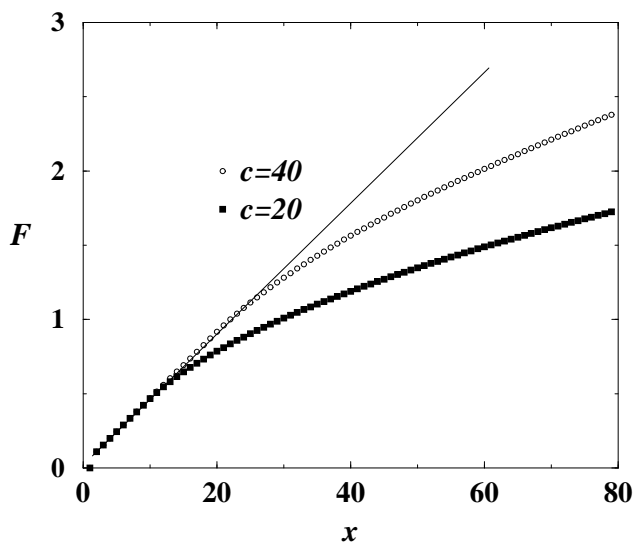}
\caption{The intermittency factor $F$
Vs. $x$, for different circumferences of the cylinder. The straight line is the 
result of two-dimensional simulations with the same spot birth rate.}
\label{fig:f2}
\end{center}
\end{figure}

\begin{figure}[h]
\begin{center}
\includegraphics[scale=0.5]{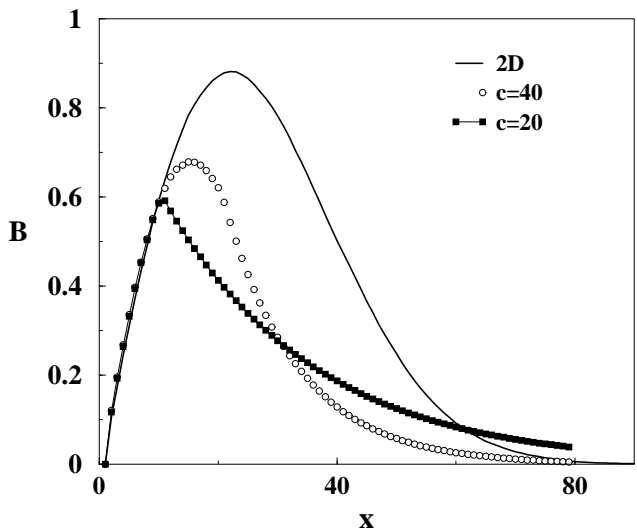}
\caption{The burst rate in the transition zone of the boundary layer
around a cylinder.}
\label{fig:burst_ax}
\end{center}
\end{figure}

We have carried out stochastic simulations 
of the birth and downstream propagation and growth of
turbulent spots in the transition zone of an axisymmetric boundary layer.
The downstream variation of the intermittency parameter $F$
is shown in figure \ref{fig:f2}. The quantity $c$ is the
circumference of the body. In the initial region, transition proceeds
exactly as it would in two-dimensional flow. This is because spots are too 
small to ``see" the body. When spots wrap themselves around the cylinder, 
a qualitative change is observed, as is expected from the discussion above,
and transition proceeds much
more slowly after this. The burst rate $B$ shown in figure \ref{fig:burst_ax}
is another indication of the differences in the transition process. 

\section{Summary and discussion}
\label{conc}

Stochastic simulations, inspired by a cellular-automaton approach,
of the generation and propagation of turbulent spots in transitional 
boundary layers have been
conducted, employing the observation that spot growth is self-similar. 
It is demonstrated that the pattern of spot birth may be inferred from the 
downstream variation of average quantities in the transition zone, such as the
intermittency and the burst rate. This is because the qualitative behaviour in
the transition zone depends on whether there is a pattern in spot merger, and 
if so, whether lateral or longitudinal merger is dominant. Contrary to present
belief, our results indicate that relatively simple experiments can tell us a 
great deal about the connection between instability and transition to 
turbulence. We show that experiments conducted in decelerating flows are much 
more conducive to exploring this connection than constant pressure flows.
Transverse curvature has the effect of slowing down the transition process. 
 
The simulations of \cite{jd01} showed a degree of spanwise
periodicity in the arrangement of the backward jets (the precursors to
spots). Only some of the jets however gave rise to turbulent spots.
It has been our contention in TS-mode driven
transition that the degree of randomness in the birth of spots
is expected to decrease as the pressure gradient becomes increasingly 
adverse. It would be interesting to see whether this is true of bypass 
transition as well. A direct numerical simulation of an adverse pressure
gradient flow should, if this contention is right, produce turbulent
spots from many more of the reverse jets.

{\bf Acknowledgments} \\
We are most grateful to Prof. Narasimha for suggesting cellular automaton
type simulations for the transition zone, and for several discussions.
Financial support from the Aeronautical R\&D Board, Government of India, is 
gratefully acknowledged.

%\bibliography{/nfs/varsha/vinod/mt/reference}

\end{document}